\documentclass[conference]{IEEEtran}

\IEEEoverridecommandlockouts

\usepackage{amsmath,amsthm,relsize}
\usepackage{amsfonts, amssymb, cuted}
\usepackage{cleveref}
\usepackage{mathtools}
\usepackage{algorithm}
\usepackage[noend]{algpseudocode}
\usepackage{cite}
\usepackage{xcolor}
\usepackage{soul}
\usepackage{graphicx}
\usepackage{tikz}
\usepackage{pgfplotstable}
\usepackage{pgfplots}
\usepackage{nicefrac}
\usepackage{enumitem}
\usepackage{support-caption}
\usepackage{subcaption}
\usepackage[font=footnotesize]{caption}
\usepackage{multirow}
\pgfplotsset{compat=1.14}
\usepackage{relsize}
\usetikzlibrary{patterns}
\usepackage{acro}
\usepackage{todonotes}

\newcommand{\vbar}{\raisebox{.17ex}{\rule{.04em}{1.35ex}}}
\newcommand{\vbarind}{\raisebox{.01ex}{\rule{.04em}{1.1ex}}}
\newcommand{\R}{\ifmmode{\rm I}\hspace{-.2em}{\rm R} \else ${\rm I}\hspace{-.2em}{\rm R}$ \fi}
\newcommand{\T}{\ifmmode{\rm I}\hspace{-.2em}{\rm T} \else ${\rm I}\hspace{-.2em}{\rm T}$ \fi}
\newcommand{\N}{\ifmmode{\rm I}\hspace{-.2em}{\rm N} \else \mbox{${\rm I}\hspace{-.2em}{\rm N}$} \fi}
\newcommand{\B}{\ifmmode{\rm I}\hspace{-.2em}{\rm B} \else \mbox{${\rm I}\hspace{-.2em}{\rm B}$} \fi}
\newcommand{\Hil}{\ifmmode{\rm I}\hspace{-.2em}{\rm H} \else \mbox{${\rm I}\hspace{-.2em}{\rm H}$} \fi}
\newcommand{\C}{\ifmmode\hspace{.2em}\vbar\hspace{-.31em}{\rm C} \else \mbox{$\hspace{.2em}\vbar\hspace{-.31em}{\rm C}$} \fi}
\newcommand{\Cind}{\ifmmode\hspace{.2em}\vbarind\hspace{-.25em}{\rm C} \else \mbox{$\hspace{.2em}\vbarind\hspace{-.25em}{\rm C}$} \fi}
\newcommand{\Q}{\ifmmode\hspace{.2em}\vbar\hspace{-.31em}{\rm Q} \else \mbox{$\hspace{.2em}\vbar\hspace{-.31em}{\rm Q}$} \fi}
\newcommand{\Z}{\ifmmode{\rm Z}\hspace{-.28em}{\rm Z} \else ${\rm Z}\hspace{-.28em}{\rm Z}$ \fi}

\DeclareAcronym{AWGN}{
    short = AWGN,
    long = additive white Gaussian noise,
    list = Additive White Gaussian Noise,
    tag = abbrev
}

\DeclareAcronym{ADMM}{
    short = ADMM,
    long = alternating direction method of multipliers,
    list = Alternating Direction Method of Multipliers,
    tag = abbrev
}

\DeclareAcronym{MGMC}{
    short = MGMC,
    long = multi-group multi-casting,
    list = multi-group multi-casting,
    tag = abbrev
}

\DeclareAcronym{SGMC}{
    short = SGMC,
    long = single-group multi-casting,
    list = single-group multi-casting,
    tag = abbrev
}
\DeclareAcronym{AoA}{
    short = AoA,
    long = angle-of-arrival,
    list = Angle-of-Arrival,
    tag = abbrev
}

\DeclareAcronym{AoD}{
    short = AoD,
    long = angle-of-departure,
    list = Angle-of-Departure,
    tag = abbrev
}

\DeclareAcronym{KKT}{
    short = KKT,
    long = Karush-Kuhn-Tucker,
    list = Karush-Kuhn-Tucker,
    tag = abbrev
}

\DeclareAcronym{MMF}{
    short = MMF,
    long = max-min-fairness,
    list = max-min-fairness,
    tag = abbrev
}

\DeclareAcronym{WMMF}{
    short = WMMF,
    long = weighted max-min-fairness,
    list = max-min-fairness,
    tag = abbrev
}

\DeclareAcronym{BB}{
    short = BB,
    long = base band,
    list = Base Band,
    tag = abbrev
}

\DeclareAcronym{BC}{
    short = BC,
    long = broadcast channel,
    list = Broadcast Channel,
    tag = abbrev
}

\DeclareAcronym{BS}{
    short = BS,
    long = base station,
    list = Base Station,
    tag = abbrev
}

\DeclareAcronym{BR}{
    short = BR,
    long = best response,
    list = Best Response, 
    tag = abbrev
}

\DeclareAcronym{CB}{
    short = CB,
    long = coordinated beamforming,
    list = Coordinated Beamforming,
    tag = abbrev
}

\DeclareAcronym{CC}{
    short = CC,
    long = coded caching,
    list = Coded Caching,
    tag = abbrev
}

\DeclareAcronym{CE}{
    short = CE,
    long = channel estimation,
    list = Channel Estimation,
    tag = abbrev
}

\DeclareAcronym{CoMP}{
    short = CoMP,
    long = coordinated multi-point transmission,
    list = Coordinated Multi-Point Transmission,
    tag = abbrev
}

\DeclareAcronym{CRAN}{
    short = C-RAN,
    long = cloud radio access network,
    list = Cloud Radio Access Network,
    tag = abbrev
}

\DeclareAcronym{CSE}{
    short = CSE,
    long = channel specific estimation,
    list = Channel Specific Estimation,
    tag = abbrev
}

\DeclareAcronym{CSI}{
    short = CSI,
    long = channel state information,
    list = Channel State Information,
    tag = abbrev
}

\DeclareAcronym{CSIT}{
    short = CSIT,
    long = channel state information at the transmitter,
    list = Channel State Information at the Transmitter,
    tag = abbrev
}

\DeclareAcronym{CU}{
    short = CU,
    long = central unit,
    list = Central Unit,
    tag = abbrev
}

\DeclareAcronym{D2D}{
    short = D2D,
    long = device-to-device,
    list = Device-to-Device,
    tag = abbrev
}

\DeclareAcronym{DE-ADMM}{
    short = DE-ADMM,
    long = direct estimation with alternating direction method of multipliers,
    list = Direct Estimation with Alternating Direction Method of Multipliers,
    tag = abbrev
}

\DeclareAcronym{DE-BR}{
    short = DE-BR,
    long = direct estimation with best response,
    list = Direct Estimation with Best Response,
    tag = abbrev
}

\DeclareAcronym{DE-SG}{
    short = DE-SG,
    long = direct estimation with stochastic gradient,
    list = Direct Estimation with Stochastic Gradient,
    tag = abbrev
}

\DeclareAcronym{DFT}{
	short = DFT,
	long = discrete fourier transform,
	list = Discrete Fourier Transform,
	tag = abbrev
}

\DeclareAcronym{DoF}{
    short = DoF,
    long = degrees of freedom,
    list = Degrees of Freedom,
    tag = abbrev
}

\DeclareAcronym{DL}{
    short = DL,
    long = downlink,
    list = Downlink,
    tag = abbrev
}

\DeclareAcronym{GD}{
	short = GD, 
	long = gradient descent,
	list = Gradeitn Descent,
	tag = abbrev
}

\DeclareAcronym{IBC}{
    short = IBC,
    long = interfering broadcast channel,
    list = Interfering Broadcast Channel,
    tag = abbrev
}

\DeclareAcronym{i.i.d.}{
    short = i.i.d.,
    long = independent and identically distributed,
    list = Independent and Identically Distributed,
    tag = abbrev
}

\DeclareAcronym{JP}{
    short = JP,
    long = joint processing,
    list = Joint Processing,
    tag = abbrev
}

\DeclareAcronym{LOS}{
	short = LOS,
	long = line-of-sight,
	list = Line-of-Sight,
	tag = abbrev
}

\DeclareAcronym{LS}{
    short = LS,
    long = least squares,
    list = Least Squares,
    tag = abbrev
}

\DeclareAcronym{LTE}{
    short = LTE,
    long = Long Term Evolution,
    tag = abbrev
}

\DeclareAcronym{LTE-A}{
    short = LTE-A,
    long = Long Term Evolution Advanced,
    tag = abbrev
}

\DeclareAcronym{MIMO}{
    short = MIMO,
    long = multiple-input multiple-output,
    list = Multiple-Input Multiple-Output,
    tag = abbrev
}

\DeclareAcronym{MISO}{
    short = MISO,
    long = multiple-input single-output,
    list = Multiple-Input Single-Output,
    tag = abbrev
}

\DeclareAcronym{MAC}{
    short = MAC,
    long = multiple access channel,
    list = Multiple Access Channel,
    tag = abbrev
}

\DeclareAcronym{MSE}{
    short = MSE,
    long = mean-squared error,
    list = Mean-Squared Error,
    tag = abbrev
}

\DeclareAcronym{MMSE}{
    short = MMSE,
    long = minimum mean-squared error,
    list = Minimum Mean-Squared Error,
    tag = abbrev
}

\DeclareAcronym{mmWave}{
	short = mmWave,
	long = millimeter wave,
	list = Millimeter Wave,
	tag = abbrev
}

\DeclareAcronym{MU-MIMO}{
    short = MU-MIMO,
    long = multi-user \ac{MIMO},
    list = Multi-User \ac{MIMO},
    tag = abbrev
}

\DeclareAcronym{OTA}{
    short = OTA,
    long = over-the-air,
    list = Over-the-Air,
    tag = abbrev
}

\DeclareAcronym{PSD}{
    short = PSD,
    long = positive semidefinite,
    list = Positive Semidefinite,
    tag = abbrev
}

\DeclareAcronym{QoS}{
	short = QoS,
	long = quality of service,
	list = Quality of Service,
	tag = abbrev
}

\DeclareAcronym{RCP}{
	short = RCP,
	long = remote central processor,
	list = Remote Central Processor,
	tag = abbrev
}

\DeclareAcronym{RRH}{
    short = RRH,
    long = remote radio head,
    list = Remote Radio Head,
    tag = abbrev
}

\DeclareAcronym{RSSI}{
    short = RSSI,
    long = received signal strength indicator,
    list = Received Signal Strength Indicator,
    tag = abbrev
}

\DeclareAcronym{RX}{
	short = RX,
	long = receiver,
	list = Receiver,
	tag = abbrev
}

\DeclareAcronym{SCA}{
    short = SCA,
    long = successive-convex-approximation,
    list = Successive-Convex-Approximation,
    tag = abbrev
}

\DeclareAcronym{SG}{
    short = SG,
    long = stochastic gradient,
    list = Stochastic Gradient,
    tag = abbrev
}

\DeclareAcronym{SIC}{
    short = SIC,
    long = successive interference cancellation,
    list = Successive Interference Cancellation,
    tag = abbrev
}

\DeclareAcronym{SNR}{
    short = SNR,
    long = signal-to-noise-ratio,
    list = Signal-to-Noise Ratio,
    tag = abbrev
}

\DeclareAcronym{SDR}{
    short = SDR,
    long = semi-definite-relaxation,
    list = semi-definite-relaxation,
    tag = abbrev
}

\DeclareAcronym{SINR}{
    short = SINR,
    long = signal-to-interference-plus-noise ratio,
    list = Signal-to-Interference-plus-Noise Ratio,
    tag = abbrev
}

\DeclareAcronym{SOCP}{
	short = SOCP, 
	long = second order cone program,
	list = Second Order Cone Program,
	tag = abbrev
}

\DeclareAcronym{SSE}{
    short = SSE,
    long = stream specific estimation,
    list = Stream Specific Estimation,
    tag = abbrev
}

\DeclareAcronym{SVD}{
	short = SVD,
	long = singular value decomposition,
	list = Singular Value Decomposition,
	tag = abbrev
}

\DeclareAcronym{TDD}{
	short = TDD,
	long = time division duplex,
	list = Time Division Duplex,
	tag = abbrev
}

\DeclareAcronym{TX}{
	short = TX,
	long = transmitter,
	list = Transmitter,
	tag = abbrev
}

\DeclareAcronym{UE}{
    short = UE,
    long = user equipment,
    list = User Equipment,
    tag = abbrev
}

\DeclareAcronym{UL}{
    short = UL,
    long = uplink,
    list = Uplink,
    tag = abbrev
}

\DeclareAcronym{ULA}{
	short = ULA,
	long = uniform linear array,
	list = Uniform Linear Array,
	tag = abbrev
}

\DeclareAcronym{UPA}{
    short = UPA,
    long = uniform planar array,
    list = Uniform Planar Array,
    tag = abbrev
}

\DeclareAcronym{WMMSE}{
    short = WMMSE,
    long = weighted minimum mean-squared error,
    list = Weighted Minimum Mean-Squared Error,
    tag = abbrev
}

\DeclareAcronym{WMSEMin}{
    short = WMSEMin,
    long = weighted sum \ac{MSE} minimization,
    list = Weighted sum \ac{MSE} Minimization,
    tag = abbrev
}

\DeclareAcronym{WBAN}{
	short = WBAN,
	long = wireless body area network,
	list = Wireless Body Area Network,
	tag = abbrev
}

\DeclareAcronym{WSRMax}{
    short = WSRMax,
    long = weighted sum rate maximization,
    list = Weighted Sum Rate Maximization,
    tag = abbrev
}

\theoremstyle{definition}

\newtheorem{exmp}{Example}%[section]

\newcommand{\CF}[0]{{\mathcal{F}}}

\newcommand{\CK}[0]{{\mathcal{K}}}

\newcommand{\CP}[0]{{\mathcal{P}}}

\newcommand{\CR}[0]{{\mathcal{R}}}

\newcommand{\CT}[0]{{\mathcal{T}}}

\newcommand{\Bh}[0]{{\mathbf{h}}}

\newcommand{\Bu}[0]{{\mathbf{u}}}

\newcommand{\Bw}[0]{{\mathbf{w}}}
\newcommand{\Bx}[0]{{\mathbf{x}}}
\newcommand{\By}[0]{{\mathbf{y}}}
\newcommand{\Bz}[0]{{\mathbf{z}}}

\newcommand{\BH}[0]{{\mathbf{H}}}
\newcommand{\BI}[0]{{\mathbf{I}}}

\newcommand{\BW}[0]{{\mathbf{W}}}

\newcommand{\Sfs}[0]{{\mathsf{s}}}

\newcommand{\SfG}[0]{{\mathsf{G}}}

\newcommand{\SfL}[0]{{\mathsf{L}}}

\newcommand{\SfX}[0]{{\mathsf{X}}}

\newcommand{\FillGray}[3]{\filldraw[gray!50](#3-1+0.1,#1-#2+0.1) rectangle (#3-0.1,#1-#2+1-0.1)}
\newcommand{\FillBlack}[3]{\filldraw[black!70](#3-1+0.1,#1-#2+0.1) rectangle (#3-0.1,#1-#2+1-0.1)}
\newcommand{\FillHatch}[3]{\fill[pattern=crosshatch, pattern color=black!25](#3-1,#1-#2)rectangle(#3,#1-#2+1)}
\newcommand{\PutText}[4]{\node[] at (#3-1+0.5,#1-#2+0.5) {\small #4}}

\newcommand{\subparagraph}{}
\usepackage{titlesec}
\titlespacing\section{3pt}{6pt plus 4pt minus 2pt}{6pt plus 2pt minus 2pt}
\titlespacing\subsection{3pt}{4pt plus 4pt minus 2pt}{4pt plus 2pt minus 2pt}
\titlespacing\subsubsection{3pt}{3pt plus 4pt minus 2pt}{0pt plus 2pt minus 3pt}

\title{Multicast Beamformer Design\\ for MIMO Coded Caching Systems
}

\begin{document}

\author{\IEEEauthorblockN{MohammadJavad Salehi, Mohammad NaseriTehrani and Antti T\"olli} \\
\IEEEauthorblockA{
    Centre for Wireless Communications, University of Oulu, 90570 Oulu, Finland \\
    \textrm{E-mail: \{firstname.lastname\}@oulu.fi}
    }
\thanks{
This work is supported by the Academy of Finland under grants no. 346208 (6G Flagship) and 343586 (CAMAIDE), and by the Finnish Research Impact Foundation under the project 3D-WIDE.}
}

\maketitle

\begin{abstract}
Coded caching (CC) techniques have been shown to be conveniently applicable in multi-input multi-output (MIMO) systems. In a $K$-user network with spatial multiplexing gains of $L$ at the transmitter and $G$ at every receiver, if each user can cache a fraction $\gamma$ of the file library, a total number of $GK\gamma + L$ data streams can be served in parallel. In this paper, we focus on improving the finite-SNR performance of MIMO-CC systems. 
%We study two MIMO-CC schemes with the same number of parallel streams but with different transmission strategies; one based on unicasting individual data terms and the other with multicasting carefully created codewords.
We first consider a MIMO-CC scheme that relies only on unicasting individual data streams, and then, introduce a decomposition strategy to design a new scheme that delivers the same data streams through multicasting of $G$ parallel codewords.
We discuss how optimized beamformers could be designed for each scheme and use numerical simulations to compare their finite-SNR performance. It is shown that while both schemes serve the same number of streams, multicasting provides notable performance improvements. This is because, with multicasting, transmission vectors are built with fewer beamformers, leading to more efficient usage of available power resources.

%We show how a proper decoding order across multiple parallel streams at the receivers allows to enhance the multicasting opportunities during the delivery phase, hence improving the achievable finite-SNR rate. We also propose a simple algorithm to construct XOR codewords and corresponding multicast beamformers of size two, effectively halving the total number of required beamformers. The simulation results demonstrate significant... .

%The abstract goes here ...

%It has been shown that in MIMO setups, the caching gain can be multiplied by the receiver-side multiplexing gain, and still be added with the transmitter-side spatial multiplexing gain. This has been possible with a pure signal-level coded caching scheme, which achieves the same DoF but ignores the CC-aided multicasting gain stemming from the XOR terms. The inferior CC-aided multicasting gain is shown to be more prominent in the finite-SNR regime, and hence, the finite-SNR performance of the proposed MIMO CC scheme would fall short of expectations. In this paper, we present a novel idea to enable XOR terms in baseline MIMO CC schemes to boost the finite-SNR performance of such schemes. We use an ordering mechanism in the receiver, to enable a hybrid bit- and signal-level CC scheme that has XOR terms while reaching the full DoF. Simulation results show noticeable performance gains, up to 50\%, in the finite-SNR regime.

\end{abstract}

\begin{IEEEkeywords}
coded caching, MIMO communications, finite-SNR performance
\end{IEEEkeywords}

\section{Introduction}
Wireless communication networks are under mounting pressure to support exponentially increasing volumes of multimedia content~\cite{cisco2018cisco} and to provide the infrastructure for the imminent emergence of new applications such as wireless immersive viewing~\cite{mahmoodi2021non,salehi2022enhancing}. 
For the efficient delivery of such multimedia content, the work of  Maddah-Ali and Niesen in~\cite{maddah2014fundamental} proposed the idea of coded caching (CC) as a means of increasing the data rates by exploiting cache content across the network.
In a single-stream downlink network of $K$ users each with a cache memory large enough to store a portion $\gamma \le 1$ of the entire file library, CC enables boosting the achievable rate by a factor of $K\gamma + 1$.
This is achieved by multicasting carefully created codewords 
%to groups of users of size $K\gamma + 1$, 
such that each user can use its cache content to remove unwanted parts from the received signal.
As a result, with CC, the achievable \ac{DoF} is increased from one to $K\gamma+1$. The multiplicative factor $t\equiv K\gamma$ is also called the \emph{coded caching gain}.

%\todo[inline]{MISO and MIMO parts could be mixed with each other, and the subpacketization part could be removed.}

Motivated by the growing importance of multi-antenna wireless communications~\cite{rajatheva2020white}, the authors in~\cite{shariatpanahi2016multi,shariatpanahi2018physical} explored the cache-aided \ac{MISO} setting, revealing that 
%the same CC gain $t$ could be achieved cumulatively with the spatial multiplexing gain. More exactly, 
for a downlink \ac{MISO} setup with cache-enabled users and the transmitter-side multiplexing gain of $L$, the cumulative \ac{DoF} of $t+L$ is achievable. Many subsequent works then explored various implementation challenges of MISO-CC schemes, including but not limited to, optimal beamformer design~\cite{tolli2017multi,tolli2018multicast,mahmoodi2021low}, exponentially growing subpacketization~\cite{lampiris2018adding,salehi2020lowcomplexity}, and applicability to dynamic setups~\cite{salehi2021low,abolpour2022coded}.

Compared to \ac{MISO} setups, applying CC techniques in \ac{MIMO} communications has been less investigated. In~\cite{cao2017fundamental,cao2019treating}, information-theoretic approaches were used to find upper bounds to the achievable \ac{DoF} in \ac{MIMO}-CC setups with multiple transmitters and receivers. It was revealed that multiple antennas at the receiver side could actually increase the \ac{DoF} over \ac{MISO} settings. Following the same direction, the authors in~\cite{salehi2021MIMO} proposed a low-complexity solution to construct \ac{MIMO}-CC schemes building on any given scheme available for \ac{MISO} setups. It was shown that for the simple case of a \ac{MIMO} setting with a single transmitter and multiple receivers with spatial multiplexing gains of $L$ and $G$, respectively, if $\frac{L}{G}$ is an integer, the cumulative \ac{DoF} of $Gt+L$ is achievable. In other words, multiple receive antennas enable a multiplicative boost in the coded caching gain.

In this paper, we study the finite-SNR performance of \ac{MIMO}-CC schemes. We start with the scheme in~\cite{salehi2021MIMO}, which is based on unicasting individual data terms, and explain how optimized beamformers could be designed for this scheme following a similar approach to~\cite{tolli2017multi}. Then, we discuss how the underlying \ac{MIMO} structure could be used to decompose the system into multiple parallel \ac{MISO} setups (or single-antenna, if $L=G$), and how this decomposition could enable transmitting several multicast codewords simultaneously. While both considered schemes deliver the same number of streams in each transmission, we expect the scheme with multicasting to have better performance in the finite-SNR regime as it requires fewer beamformers in each transmission (hence, the usage of available power resources is more efficient). Finally, numerical simulations are used to verify this assumption.

Throughout the text, we use the following notations. For integer $J$, $[J]$ is the set of numbers $\{1,2,\cdot \cdot \cdot,J\}$. 
Boldface upper- and lower-case letters indicate matrices and vectors, respectively, and
%and the set of users is shown by $\CK$. A general file in the library is shown by $W$ and the file requested by user $u_k \in \CK$ is denoted by $W(k)$. 
%Zero-forcing transmit beamformer vectors are denoted by $\Bw_{\CR}$, where the set $\CR$ clarifies the zero-forcing locations. For simplicity, we remove the brackets while explicitly writing indices in $\CR$. We use $\Sfs_k^g$, $g \in [G]$, to refer to the data stream $g$ intended for user $u_k$. %Finally, the bit-wise XOR operation is denoted by $\oplus$.
calligraphic letters denote sets. 
%$\BC[i,j]$ is the element in row $i$ and column $j$ of matrix $\BC$, and $[\BA;\BB]$ represents the matrix formed by horizontal concatenation of two matrices $\BA$ and $\BB$.
For two sets $\CK$ and $\CT$, $\CK \backslash \CT$ is the set of elements in $\CK$ that are not in $\CT$.
%For two sets $\CA$ and $\CB$, $\CA \backslash \CB$ denotes the set of elements in $\CA$ that are not in $\CB$. 
Other notations are defined as they are used in the text.

\section{System Model}
\label{section:sys_model}
%\subsection{Network Setup}
We consider a MIMO communication setup with $K$ users, $\SfL$ antennas at the transmitter, and $\SfG$ antennas at each receiver. We assume the spatial multiplexing gains at the transmitter and receivers are set to $L \le \SfL$ and $G \le \SfG$, i.e., the transmitter can deliver $L$ data streams and each user can receive $G$ data streams simultaneously.
This requires that the channel matrices $\BH_k \in \mathbb{C}^{\SfG \times \SfL}$ from the transmitter to every user $k \in [K]$ to have ranks not smaller than $G$,
and the cumulative channel matrix formed by the vertical concatenation of individual channel matrices as
%defined by the vertical concatenation of individual channel matrices as
$\BH = [\BH_1^H; \BH_2^H; \cdots; \BH_k^H]^H$ to have a rank larger than or equal to $L$.
%and a single server, where the server has $\SfL$ transmit antennas and every user has $\SfG$ receive antennas. 
%These antenna arrays enable spatial multiplexing gains of $L \le \SfL$ and $G \le \SfG$ to be achievable at the transmitter and receiver side, respectively. In other words, if the $\SfG \times \SfL$ channel matrix from the transmitter to user $k \in [K]$ is shown by $\BH_k$, the rank of every $\BH_k$ is at least $G$, and the rank of the cumulative channel matrix, defined by the vertical concatenation of individual channel matrices as
%\begin{equation}
%\BH =
%\begin{bmatrix}
%    \BH_1 \\
%    \cdot \cdot \cdot \\
%    \BH_K
%\end{bmatrix}
%\; 
%\end{equation}
%to have ranks larger than or equal to $L$ and $G$, respectively.
%is at least $L$. 
%This assumption enables the transmitter to deliver $L$ independent data streams and each user to receive $G$ independent streams simultaneously. 
%
Each user has a cache memory of size $MF$ bits, and requests files from a library $\CF$ of $N$ files each with size $F$ bits. For notational simplicity, we use a normalized data unit and ignore $F$ in subsequent notations. The coded caching gain is defined as $t \equiv \frac{KM}{N}$ and indicates
%The value of $t$ also indicates 
how many copies of the file library can be stored in the cache memories of all the users. 
We assume both $\eta \equiv \frac{L}{G}$ and $t$ are integers.
%is an integer and leave the analysis for the non-integer case to the extended version.

The system operation consists of two phases; placement, and delivery. In the placement phase, users' cache memories are filled up with data. Following a similar structure as in~\cite{shariatpanahi2018physical}, we split each file $W \in \CF$ into $\binom{K}{t}$ subfiles $W_{\CP}$, where $\CP \subseteq [K]$ can be any subset of users with $|\CP| = t$. Then, in the cache memory of user $k \in [K]$, we store $W_{\CP}$, $\forall W \in \CF$, $\forall \CP : k \in \CP$.

At the beginning of the delivery phase, every user $k$ announces its requested file $W(k) \in \CF$ to the server. The server then builds and transmits a vector $\Bx(\CK)$ for every subset of users $\CK \subseteq [K]$ with $|\CK| = t+\eta$ (transmissions are done, e.g., in consecutive TDMA slots). Each $\Bx(\CK)$ is built to deliver $G$ parallel data streams to every user in $\CK$, resulting in the total \ac{DoF} of $G(t+\eta) = Gt+L$.
After transmitting $\Bx(\CK)$, every user $k \in \CK$ receives $\By_k(\CK) = \BH_k \Bx(\CK) + \Bz_k$, where $\Bz_k \in \mathbb{C}^{\SfG \times 1}$ 
is the additive white Gaussian noise (AWGN) with power $N_0$.
%represents the noise vector. 
Then, receive beamforming vectors $\Bu_{k,g} \in \mathbb{C}^{\SfG \times 1}$, $g \in [G]$ are used to produce stream-specific received signals $y_{k,g}(\CK) = \Bu_{k,g}^H \By_k(\CK)$. For simplicity, let us use $\Sfs_{k,g}$ to denote the stream $g$ at user $k$, and define the equivalent channel vector for the stream $\Sfs_{k,g}$ as $\Bh_{k,g} = (\Bu_{k,g}^H \BH_{k})^H $. Then, we have
\begin{equation}
\label{eq:received_signal_model}
    y_{k,g}(\CK) = \Bh_{k,g}^H \Bx(\CK) + z_{k,g} \; ,
\end{equation}
where $z_{k,g} = \Bu_{k,g}^H \Bz_k$ denotes the AWGN over the stream $\Sfs_{k,g}$.

The transmission vectors $\Bx(\CK)$ are built using the delivery algorithm of the considered \ac{MIMO}-CC scheme. In this paper, we study two schemes with the same number of parallel data streams but with different transmission strategies; unicasting and multicasting. The details of the schemes are provided in subsequent sections.

\section{Data Delivery with Unicasting}
\label{section:unicast_delivery}
For data delivery with unicasting, we consider the \ac{MISO} scheme in~\cite{shariatpanahi2018physical} as the baseline and use the stretching mechanism in~\cite{salehi2021MIMO} to apply it to \ac{MIMO} setups. 
As a result, we first split every subfile $W_{\CP}$ into $Q = \binom{K-t-1}{L-1}$ smaller parts $W_{\CP}^q$, $q \in [Q]$, and then, split every resulting part $W_{\CP}^q$ into $G$ subpackets $W_{\CP}^{q,g}$, $g \in [G]$. The index $q$ does not affect our analysis in this paper and is removed for notational simplicity.
Then, the transmission vector $\Bx(\CK)$ is built as
\begin{equation}
\label{eq:trans_vector_general}
    \Bx(\CK) = \sum_{\substack{\CT \subseteq \CK \\ |\CT|=t+1}} \sum_{\substack{k \in \CT}} \sum_{G \in [G]} W_{\CT \backslash \{k\}}^{g} (k) \Bw_{\CR(\CK,\CT,k,g)} \; ,
\end{equation}
%where $\CT \backslash k$ denotes the set of users in $\CT$ except $k$, and 
where 
%$q$ is the second-level subpacketization index used to ensure fresh data bits are sent in each transmission (c.f.~\cite{shariatpanahi2018physical}). The $q$ index does not affect our analysis in this paper and is removed for notational simplicity. Moreover, 
%the set %$\CR(\cdot)$ represents the set of streams over which the interference should be suppressed by beamforming and is built as
\begin{equation}
    \CR(\CK,\CT,k,g) = \bigcup_{\substack{\Bar{k} \in \CK \backslash \CT \\ \Bar{g} \in [G]}} \{ \Sfs_{\Bar{k},\Bar{g}}\} \; \bigcup_{\substack{\Bar{g} \in [G]\backslash \{g\} }} \{ \Sfs_{k,\Bar{g}} \}
\end{equation}
represents the set of streams over which the interference is suppressed by beamforming, and $\Bw_{\CR(\cdot)}$ denotes transmit beamformers.
As a quick explanation, using vector $\Bx(\CK)$, we transmit $G(t+1)\binom{t+\eta}{t+1}$ subpackets (recall that $|\CK| = t+\eta$), and all these subpackets can be decoded simultaneously as the interference caused by each of them over other streams is either suppressed by beamforming or could be reconstructed and removed using the cache contents.
%Also, the interference caused by the subpacket $W_{\CT \backslash \{k\}}^{g} (k)$ is:
%\begin{itemize}
%    \item reconstructed and removed by every user $\Bar{k} \in \CT \backslash \{k\}$, as these users have the subpacket in their cache,% memory, and
%    \item suppressed by beamforming 
%\end{itemize}
The following example clarifies this procedure.

%In this paper, we use the same successive convex approximation technique in~\cite{tolli2017multi} and alternating optimization to design optimized transmit and receive beamformers.

%The following example clarifies how the transmission vectors are built for a small \ac{MIMO} setup.
%Without loss of generality, we assume that $\Bx(\CK)$ vectors are sent in a TDMA manner, i.e., in consecutive time intervals.

\begin{exmp}
    \label{exmp:wsa_k2}
    Consider a \ac{MIMO} setup where spatial multiplexing gains at the transmitter and receivers are $L=G=2$, and the coded caching gain is $t=1$ (i.e., $t+\eta = 2$). %Following the \ac{MIMO}-CC scheme in~\cite{salehi2021MIMO}, we can serve $t + \eta = 2$ users in each transmission. 
    Let us consider the transmission vector serving users~1 and~2, assuming they have requested files $A$ and $B$, respectively. It is built as
    \begin{equation}
    \label{eq:trans_vector_wsa_exmp}
        \Bx(\{1,2\}) = A_2^1 \Bw_{\Sfs_{1,2}} + A_2^2 \Bw_{\Sfs_{1,1}} + B_1^1 \Bw_{\Sfs_{2,2}} + B_1^2 \Bw_{\Sfs_{2,1}} \; , 
    \end{equation}
    and delivers four subpackets to users~1 and~2 in parallel (note that the brackets for sets $\CR(\cdot)$ are removed for notational simplicity).
    Let us review the decoding process for the first subpacket for user~1, i.e., $A_2^1$. Using~\eqref{eq:received_signal_model}, the
    respective stream-specific received signal for this subpacket is
    %received signal for this data stream is
    \begin{equation*}
    \begin{aligned}
        y_{1,1}(\{1,2\}) &= \Bh_{1,1}^H \Bx(\{1,2\}) + z_{1,1} \\
        &= A_2^1 \Bh_{1,1}^H\Bw_{\Sfs_{1,2}} + A_2^2 \Bh_{1,1}^H\Bw_{\Sfs_{1,1}} + I_c + z_{1,1} ,
    \end{aligned}
    \end{equation*}
    where the interference term $I_c = B_1^1 \Bh_{1,1}^H\Bw_{\Sfs_{2,2}} + B_1^2 \Bh_{1,1}^H\Bw_{\Sfs_{2,1}}$ can be fully reconstructed and removed using the cache contents of user~1.\footnote{Similar to~\cite{salehi2021MIMO}, we assume that equivalent channel multipliers $\Bh_{k,g}^H \Bw_{\CR(\cdot)}$ could be estimated at the receivers, e.g., using downlink precoded pilots.} Now, by definition, the remaining interference term $A_2^2 \Bh_{1,1}^H\Bw_{\Sfs_{1,1}}$ is also suppressed by beamforming, and hence, $A_2^1$ is decodable at user~1. Following the same process, all the other three streams could also be decoded successfully.
\end{exmp}

In~\cite{salehi2021MIMO}, zero-force (ZF) unicast beamformers are used to completely null out the interference, i.e., transmit and receive beamformers are built such that $\Bh_{k,g}^H \Bw_{\CR(\CK,\CT,k,g)} = 0$ for every $\Sfs_{k,g} \in \CR(\CK,\CT,k,g)$. Of course, ZF beamforming is not an appropriate choice in finite-SNR~\cite{tolli2017multi}, and we need to design optimized beamformers to improve the performance in this regime. This can be done
%
%For the considered \ac{MIMO}-CC scheme, optimized beamformers could be designed 
using alternating optimization together with the successive convex approximation (SCA) method in~\cite{tolli2017multi}. To do so, at each step, we fix either transmit or receive beamformers to their latest known values and find the optimal solution for the other one, and update this procedure until the convergence is achieved.

Here, due to lack of space, we review the beamformer design process only for the symmetric case of $L=G$ (i.e., $\eta=1$) and leave the general formulation for the extended version of this paper. The reason for this assumption is that if $\eta > 1$, each receiver has to decode multiple subpackets over each stream through an equivalent user-specific multiple access channel (MAC)~\cite{tolli2017multi}, and hence, linear MMSE receivers would no longer work. However, if $\eta = 1$, after solving for optimal transmit beamformers, receive beamformers can be simply calculated as
\begin{equation}
\label{eq:lin_mmse_receiver}
    \Bu_{k,g} = (\BH_k \BW_k \BW_k^H \BH_k^H + N_0 \BI)^{-1} \BH_k \Bw_{\CR(\CK,\CK,k,g)} \; ,
\end{equation}
where $\BW_k \equiv [\Bw_{\CR(\CK,\CK,k,1)};...;\Bw_{\CR(\CK,\CK,k,G)}]$ is formed by concatenating all the $G$ transmit beamformers used to send data to user $k$ (note that when $\eta = 1$, $\CT = \CK$).

Now, assuming receive beamformers are fixed, to calculate optimal transmit beamformers we first notice that the SINR for decoding subpacket $W_{\CT \backslash \{k\}}^{g} (k)$ at user $k \in \CK$ is given as
\begin{equation}
\label{eq:sinr_general_wsa}
    \lambda_{k,g} = \frac{ |\Bh_{k,g}^H \Bw_{\CR(\CK,\CK,k,g)}|^2 }{\sum_{\substack{\Bar{g} \in [G] \backslash \{g\} }} |\Bh_{k,g}^H \Bw_{\CR(\CK,\CK,k,\Bar{g})}|^2  + N_0 }
\end{equation}
and the rate optimization problem will be
\begin{equation}
\label{eq:opt_prob_wsa}
\begin{aligned}
    &\max_{\Bw_{\CR(\CK,\CK,k,g)}} \min_{k\in \CK, g\in[G]} r_{k,g} \\
    &s.t. \quad r_{k,g} \le \log (1 + \lambda_{k,g}), \; \forall k \in \CK, g \in [G] \; , \\
    & \quad \quad \sum_{k \in \CK} \sum_{g \in [G]} |\Bw_{{\CR(\CK,\CK,k,g)}}|^2 \le P_T \; ,
\end{aligned}
\end{equation}
where $P_T$ is the total available transmit power. As discussed in~\cite{tolli2017multi}, this optimization problem is non-convex and can be solved, e.g. using an SCA method similar to~\cite{tolli2017multi}.

\section{Data Delivery with Multicasting}
\label{section:multicast_delivery}
The primary goal of using multicast transmissions in the \ac{MIMO}-CC scheme is to improve the finite-SNR performance. This is because, with multicasting, fewer beamforming vectors are needed for the transmission, and as a result, the average power allocated to each beamformer is increased. Of course, this will have a more prominent effect in finite-SNR communications as the rate is power-limited in this regime. Similar results for \ac{MISO} setups could be found in~\cite{salehi2019subpacketization,salehi2022multi}.

\begin{figure}[t]
    %\begin{subfigure}{0.3\columnwidth}
        \centering
        \includegraphics[height = 5cm]{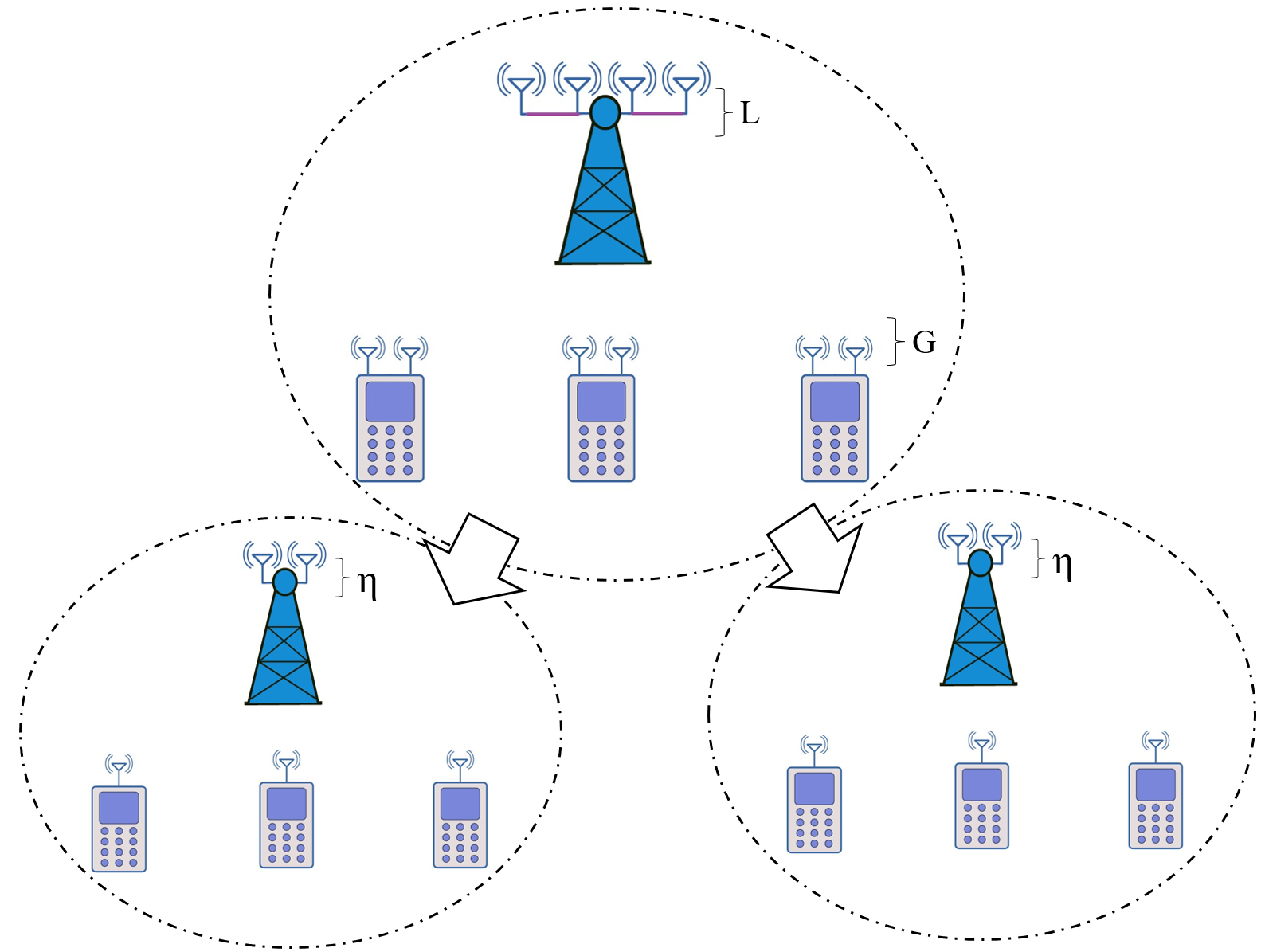}
        %\vspace{5pt}
     %   \caption{WSA}
     %   \label{fig:receiver_WSA}
    %\end{subfigure}
    %\hspace{10pt}
    %\begin{subfigure}{0.7\columnwidth}
    %    \centering
    %    \includegraphics[height = 5cm]{Figs/Receiver_Prop_2.png}
    %    \caption{Proposed}
    %    \label{fig:receiver_Prop}
    %\end{subfigure}
    %\vspace{-10pt}
    \caption{MIMO system decomposition into parallel MISO systems}
    \label{fig:decompose}
\end{figure}

In order to illustrate the multicasting opportunities available in MIMO-CC, we decompose the \ac{MIMO} system into $G$ parallel \ac{MISO} systems (or single-antenna, if $L=G$), as shown in Figure~\ref{fig:decompose}. This requires splitting subfiles $W_{\CP}$ into subpackets $W_{\CP}^{q,g}$ following the same process as in Section~\ref{section:unicast_delivery}, and building the multicast-enabled transmission vector $\hat{\Bx}(\CK)$ as
\begin{equation}
    \hat{\Bx}(\CK) = \sum_{\substack{\CT \subseteq \CK \\ |\CT|=t+1}} \sum_{\substack{g \in [G]}} \SfX_{\CT}^{g} \hat{\Bw}_{\hat{\CR}(\CK,\CT,g)} \; ,
\end{equation}
where, using $\oplus$ to represent the bit-wise XOR operation in the finite field, we have %the codeword $\SfX_{\CT}^{g}$ is built as
\begin{align}
    &\SfX_{\CT }^{g} = \bigoplus_{k \in \CT} W_{\CT \backslash \{k\}}^{g} (k) \; , \\
    &\hat{\CR}(\CK,\CT,g) = \bigcup_{\substack{\Bar{k} \in \CK \backslash \CT \\ \Bar{g} \in [G]}} \{ \Sfs_{\Bar{k},\Bar{g}}\} \; \bigcup_{\substack{\Bar{k} \in \CT \\ \Bar{g} \in [G] \backslash \{g\}}} \{ \Sfs_{\Bar{k},\Bar{g}} \} \; .
\end{align}
Note that the $q$ index is again removed for notational simplicity.
As a quick explanation, instead of unicasting individual subpackets, here we deliver $G$ parallel codewords to every subset of users $\CT \subseteq \CK$ with size $|\CT| = t+1$. Each codeword is built using the same XOR method as in~\cite{shariatpanahi2018physical} and includes fresh subpackets for all the $t+1$ users in $\CT$. Every transmission vector $\hat{\Bx}(\CK)$ delivers $G\binom{t+\eta}{t+1}$ codewords, and hence, the same number of $G(t+1)\binom{t+\eta}{t+1}$ subpackets as the unicast-based scheme in Section~\ref{section:unicast_delivery}.
The following example clarifies data delivery with the proposed scheme.

\begin{exmp}
    \label{exmp:proposed_k2}
    Consider the network in Example~\ref{exmp:wsa_k2}.
    Instead of using unicast transmission as in~\eqref{eq:trans_vector_wsa_exmp}, we can transmit two codewords $\SfX_{\{1,2\}}^1 = A_2^1 \oplus B_1^1$ and $\SfX_{\{1,2\}}^2 = A_2^2 \oplus B_1^2$ using
    \begin{equation}
        \hat{\Bx}(\{1,2\}) = \SfX_{\{1,2\}}^1\hat{\Bw}_{\Sfs_{1,2},\Sfs_{2,2}} + \SfX_{\{1,2\}}^2\hat{\Bw}_{\Sfs_{1,1},\Sfs_{2,1}} .
    \end{equation}
    Then, for decoding $A_2^1$ at user~1, this user has to first extract $\SfX_{\{1,2\}}^1$ from the stream-specific received signal
    \begin{equation*}
        \hat{y}_{1,1}(\{1,2\}) \! = \SfX_{\{1,2\}}^1 \Bh_{1,1}^H \hat{\Bw}_{\Sfs_{1,2},\Sfs_{2,2}} + \SfX_{\{1,2\}}^2 \Bh_{1,1}^H \hat{\Bw}_{\Sfs_{1,1},\Sfs_{2,1}} + z_{1,1},
    \end{equation*}
    which is possible as the interference term $\SfX_{\{1,2\}}^2 \Bh_{1,1}^H \hat{\Bw}_{\Sfs_{1,1},\Sfs_{2,1}}$ is suppressed by beamforming. Finally, after extracting $\SfX_{\{1,2\}}^1$, user~1 can decode $A_2^1$ by removing $B_1^1$ which is available in its cache memory. Using the same procedure, all four subpackets could be decoded successfully.
\end{exmp}

We can use a similar procedure as the unicast case to design optimized beamformers for the multicast-enabled scheme. Again, due to lack of space, we only review this process for the symmetric case of $L=G$ (and hence, $\CT = \CK$). With this assumption, updating received beamformers is possible using the same linear MMSE receiver in~\ref{eq:lin_mmse_receiver}. However, to find optimal transmit beamformers, instead of using the SINR value in~\eqref{eq:sinr_general_wsa} for decoding subpackets, we need to use the SINR term for extracting codeword $\SfX_{\CK}^g$ at user $k$, given as
\begin{equation}
\label{eq:sinr_multicast}
    \hat{\lambda}_{k,g} = \frac{ |\Bh_{k,g}^H \hat{\Bw}_{\hat{\CR}(\CK,\CK,g)}|^2 }{ \sum_{\substack{\Bar{g} \in [G] \backslash \{g\} }} |\Bh_{k,g}^H \hat{\Bw}_{\hat{\CR}(\CK,\CK,\Bar{g})}|^2  + N_0 } \; .
\end{equation}
Finally, the rate optimization problem could be written as
\begin{equation}
\label{eq:opt_prob_multicast}
\begin{aligned}
    &\max_{\hat{\Bw}_{\hat{\CR}(\CK,\CK,g)}} \min_{k\in \CK, g\in[G]} \hat{r}_{k,g} \\
    &s.t. \quad \hat{r}_{k,g} \le \log (1 + \hat{\lambda}_{k,g}), \; \forall k \in \CK, g \in [G] \; , \\
    & \quad \quad \sum_{g \in [G]} |\hat{\Bw}_{{\hat{\CR}(\CK,\CK,g)}}|^2 \le P_T \; .
\end{aligned}
\end{equation}
Comparing~\eqref{eq:opt_prob_multicast} with~\eqref{eq:opt_prob_wsa}, it is clear that the number of required beamformers for every transmission is reduced by a factor of $|\CK| = t+1$. As a result, we expect a performance boost as the average power allocated to each beamformer is increased. On the other hand, from~\eqref{eq:sinr_multicast}  and~\eqref{eq:sinr_general_wsa}, it can be seen that every beamformer appears more times at the interference sum (denominators of the SINR terms), by the same factor of $t+1$. This results in more constraints as we search for the optimal beamformer, hindering the performance improvement expected by the power gain. However, as will be shown by simulation results, the overall performance improvement is still noticeable -- especially in the finite-SNR regime.

\section{Simulation Results}
We use numerical simulations to compare the performance of the proposed schemes. For reference, we also simulate another setup, called \emph{virtual \ac{MISO}}, where the spatial multiplexing gain at the receiver side is used only for achieving a beamforming gain. For this setup, we consider just one receive beamforming vector $\Bu_k$ for user~$k$, and set it as the eigenvector corresponding to the strongest eigenmode (i.e., the largest eigenvalue) of $\BH_k$ (the channel matrix of user~$k$). The coded caching scheme is also set to be the \ac{MISO} scheme in~\cite{shariatpanahi2018physical}. The goal of simulating this virtual MISO setup is to analyze the real performance gains achieved by using the proposed \ac{MIMO}-CC schemes. All the simulations are done for a network of $K=8$ users with coded caching gain $t=1$, and optimized beamformers are used for transmissions.

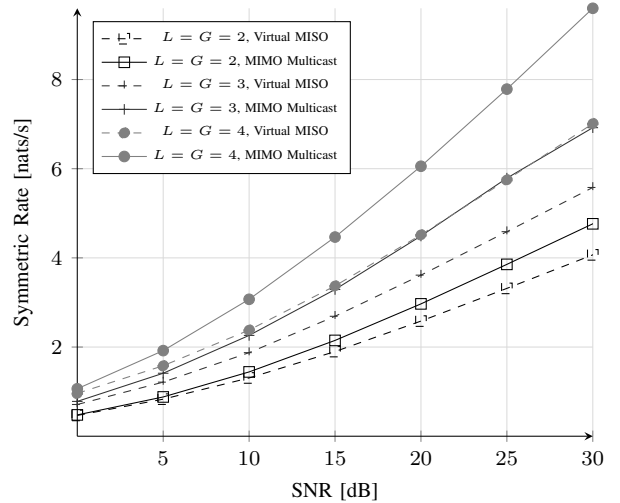
\begin{figure}
    \centering
    %\resizebox{0.9\columnwidth}{!}{%

    \begin{tikzpicture}

    \begin{axis}
    [
    % put axis lines at left and bottom
    axis lines = center,
    % control axis labels
    xlabel near ticks,
    xlabel = \smaller {SNR [dB]},
    ylabel = \smaller {Symmetric Rate [nats/s]},
    ylabel near ticks,
    ymin = 0,
    xmax = 30,
    % control legend position
    legend pos = north west,
    % control size of tick marks (10,20,30,etc)
    ticklabel style={font=\smaller},
    % control major grids
    grid=both,
    major grid style={line width=.2pt,draw=gray!30},
    % control minor grids
    %grid style={line width=.1pt, draw=gray!10},
    %minor tick num=5,
    ]
    
    %\addplot
    %[dashed, black]
    %table[y=MS-A2,x=SNR]{Data/SN1-3-Rate.tex};
    %\addlegendentry{\tiny MS-A2}
    
    \addplot
    [dashed, mark = square, black]
    table[y=LG2Virt,x=SNR]{Figs/data.tex};
    \addlegendentry{\tiny $L=G=2$, Virtual MISO}
    
    \addplot
    [mark = square, black]
    table[y=LG2Prop,x=SNR]{Figs/data.tex};
    \addlegendentry{\tiny $L=G=2$, MIMO Multicast}
    
    \addplot
    [dashed, mark = +, black!80]
    table[y=LG3Virt,x=SNR]{Figs/data.tex};
    \addlegendentry{\tiny $L=G=3$, Virtual MISO}
    
    \addplot
    [mark = +, black!80]
    table[y=LG3Prop,x=SNR]{Figs/data.tex};
    \addlegendentry{\tiny $L=G=3$, MIMO Multicast}

    \addplot
    [dashed, mark = *, black!50]
    table[y=LG4Virt,x=SNR]{Figs/data.tex};
    \addlegendentry{\tiny $L=G=4$, Virtual MISO}
    
    \addplot
    [mark = *, black!50]
    table[y=LG4Prop,x=SNR]{Figs/data.tex};
    \addlegendentry{\tiny $L=G=4$, MIMO Multicast}
    
    \end{axis}

    \end{tikzpicture}
    %}

    \caption{MIMO multicast vs Virtual MISO - $K=8$, $t=1$}
    \label{fig:plot_1}
\end{figure}

In Figure~\ref{fig:plot_1}, we have compared the performance of the virtual MISO setup with the proposed \ac{MIMO}-CC scheme with underlying multicast transmissions. As can be seen, the MIMO scheme outperforms the virtual MISO setup for all values of $L$ and $G$. Moreover, the gap between the performance of the two schemes becomes larger at higher SNR values and also as $L$ and $G$ grow. This happens because the main advantage of \ac{MIMO}-CC schemes is their larger DoF, where the coded caching gain $t$ is multiplied by $G$, and the DoF value has a more prominent performance effect in the high-SNR regime~\cite{tolli2017multi,salehi2022multi}.

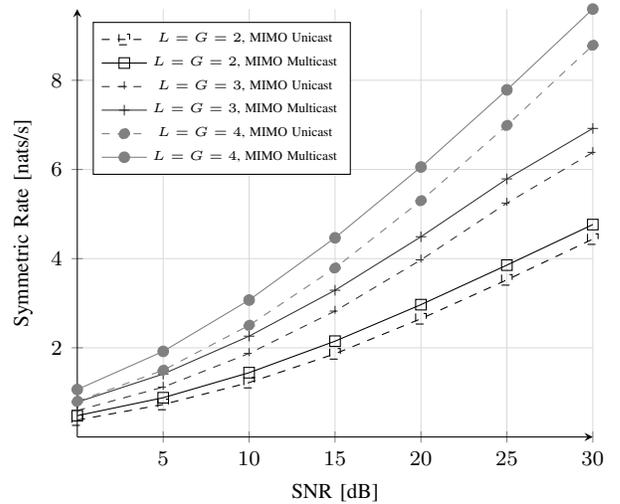
\begin{figure}
    \centering
    %\resizebox{0.9\columnwidth}{!}{%

    \begin{tikzpicture}

    \begin{axis}
    [
    % put axis lines at left and bottom
    axis lines = center,
    % control axis labels
    xlabel near ticks,
    xlabel = \smaller {SNR [dB]},
    ylabel = \smaller {Symmetric Rate [nats/s]},
    ylabel near ticks,
    ymin = 0,
    xmax = 30,
    % control legend position
    legend pos = north west,
    % control size of tick marks (10,20,30,etc)
    ticklabel style={font=\smaller},
    % control major grids
    grid=both,
    major grid style={line width=.2pt,draw=gray!30},
    % control minor grids
    %grid style={line width=.1pt, draw=gray!10},
    %minor tick num=5,
    ]
    
    %\addplot
    %[dashed, black]
    %table[y=MS-A2,x=SNR]{Data/SN1-3-Rate.tex};
    %\addlegendentry{\tiny MS-A2}
    
    \addplot
    [dashed, mark = square, black]
    table[y=LG2WSA,x=SNR]{Figs/data.tex};
    \addlegendentry{\tiny $L=G=2$, MIMO Unicast}
    
    \addplot
    [mark = square, black]
    table[y=LG2Prop,x=SNR]{Figs/data.tex};
    \addlegendentry{\tiny $L=G=2$, MIMO Multicast}
    
    \addplot
    [dashed, mark = +, black!80]
    table[y=LG3WSA,x=SNR]{Figs/data.tex};
    \addlegendentry{\tiny $L=G=3$, MIMO Unicast}
    
    \addplot
    [mark = +, black!80]
    table[y=LG3Prop,x=SNR]{Figs/data.tex};
    \addlegendentry{\tiny $L=G=3$, MIMO Multicast}

    \addplot
    [dashed, mark = *, black!50]
    table[y=LG4WSA,x=SNR]{Figs/data.tex};
    \addlegendentry{\tiny $L=G=4$, MIMO Unicast}
    
    \addplot
    [mark = *, black!50]
    table[y=LG4Prop,x=SNR]{Figs/data.tex};
    \addlegendentry{\tiny $L=G=4$, MIMO Multicast}
    
    \end{axis}

    \end{tikzpicture}
    %}

    \caption{MIMO multicast vs MIMO unicast - $K=8$, $t=1$}
    \label{fig:plot_2}
\end{figure}

In Figure~\ref{fig:plot_2}, we have compared the performance of the two MIMO-CC schemes considered in this paper. One scheme is based on unicasting subpackets and the other one incorporates multicasting codewords. As can be seen, the scheme with multicasting provides superior performance. This is expected as with multicasting, fewer beamformers are needed for transmission, and the average power allocated to each beamformer increases. Moreover, the (relative) improvement is more prominent in finite-SNR (e.g., up to 30\% for the considered network with $L=G=4$ at $5\textrm{dB}$), which is also expected as the rate is power-limited in this regime.

\section{Conclusion and Future Work}
Coded caching techniques are conveniently applicable in MIMO systems. With coded caching gain $t$, if the spatial multiplexing gains of the transmitter and receivers are $L$ and $G$, respectively, the total DoF of $Gt + L$ is achievable. In this paper, we focused on improving the finite-SNR performance of coded caching schemes for MIMO systems. We studied two schemes with the same number of parallel streams (i.e., with the same DoF) but with different transmission strategies; one based on unicasting individual data terms and the other with multicasting carefully created codewords. We discussed how optimized beamformers could be designed for each scheme and used numerical simulations to compare their finite-SNR performance. It was shown that while both schemes have the same DoF, with multicasting, the performance could be improved noticeably, especially in the finite-SNR regime. This was related to the fact that with multicasting, transmission vectors are built with fewer beamformers, and hence, the average power allocated to each beamformer is increased.

Future extensions include applying results to the non-integer $\eta$ case and the non-symmetric ($L \neq G$) scenario. We also target designing MIMO-CC schemes with multicasting but without requiring the complex successive interference cancellation (SIC) structure at the receiver.

%In this paper, we investigated how coded caching schemes can be applied to multi-input multi-output (MIMO) communication setups. We introduced a novel mechanism to elevate any multi-input single-output (MISO) scheme in the literature to be applicable to MIMO settings. The elevation mechanism keeps the properties of the baseline scheme (e.g., the low subpacketization requirement) while achieving an increased coded caching gain. We also used results from the shared-cache model to claim the optimality of the achievable degrees of freedom (DoF) of the resulting scheme among the class of linear, single-shot caching schemes with uncoded data placement. Overall, the proposed elevation mechanism allows designing efficient, low-complexity coded caching schemes for any MIMO network.

%The discussions in this paper were focused on communications in the high-SNR regime where the DoF metric is an appropriate performance indicator. However, a thorough analysis of the performance at the finite-SNR regime is due for later work. Also, investigating the applicability of the proposed scheme under dynamic network conditions where users can freely join/leave the network is part of the ongoing research.

%\section*{Extended Abstract}
%\input{ExtAbstract}

\bibliographystyle{IEEEtran}
\bibliography{references}

\end{document}